# Attention-GAN for Anomaly Detection: A Cutting-Edge Approach to Cybersecurity Threat Management

Mohammed Abo Sen [1]

*1 Grand Canyon University, 2312 Wyoming Dr, Rockford IL 61108, USA*

**Abstract**

This paper proposes an innovative Attention-GAN framework for enhancing cybersecurity, focusing on anomaly detection. In response to the challenges posed by the constantly evolving nature of cyber threats, the proposed approach aims to generate diverse and realistic synthetic attack scenarios, thereby enriching the dataset and improving threat identification. Integrating attention mechanisms with Generative Adversarial Networks (GANs) is a key feature of the proposed method. The attention mechanism enhances the model's ability to focus on relevant features, essential for detecting subtle and complex attack patterns. In addition, GANs address the issue of data scarcity by generating additional varied attack data, encompassing known and emerging threats. This dual approach ensures that the system remains relevant and effective against the continuously evolving cyberattacks. The KDD Cup and CICIDS2017 datasets were used to validate this model, which exhibited significant improvements in anomaly detection. It achieved an accuracy of 99.69% on the KDD dataset and 97.93% on the CICIDS2017 dataset, with precision, recall, and F1-scores above 97%, demonstrating its effectiveness in recognizing complex attack patterns. This study contributes significantly to cybersecurity by providing a scalable and adaptable solution for anomaly detection in the face of sophisticated and dynamic cyber threats. The exploration of GANs for data augmentation highlights a promising direction for future research, particularly in situations where data limitations restrict the development of cybersecurity systems. The attention-GAN framework has emerged as a pioneering approach, setting a new benchmark for advanced cyber-defense strategies.

## Introduction

In the digital transformation era, cybersecurity has risen from a secondary concern in information technology to a critical safeguard for digital assets and the integrity of data-centric societies [1]. This shift is largely due to increasing reliance on digital technologies, making protecting digital infrastructure more crucial than ever. The role of cybersecurity in safeguarding sensitive information, maintaining privacy, and ensuring the uninterrupted operation of critical systems is indisputable. The advent of Artificial Intelligence (AI) has introduced new dimensions in this field, providing enhanced tools for detecting and responding to cyber threats [2]. However, integrating AI into a complex cybersecurity landscape presents a multifaceted challenge.

The cyber threat landscape continuously evolves, with threats becoming increasingly sophisticated and damaging. Cybersecurity breaches result in financial losses, damage reputations, erode customer trust, and severely affect national security and public safety. As cyber criminals employ more advanced techniques, traditional security protocols that rely on static and predefined rules struggle to keep pace [3]. In this context, AI technologies, known for their capacity to learn and adapt, have emerged as a ray of hope. They offer the potential to counter these evolving threats but also bring to the forefront issues around their effectiveness, ethical implications, and operational transparency [4].

The current cybersecurity environment is a dynamic and evolving battlefield characterized by the emergence of increasingly sophisticated and complex threats. As a result, the frequency and severity of cyber incidents continue to increase, posing significant challenges to traditional security measures. These measures, largely based on historical data and known threat patterns, are becoming less effective when faced with rapidly changing threats. They are ill-equipped to handle novel or advanced cyber-attacks, which can result in significant financial losses, critical service disruptions, theft of intellectual property, and breaches of sensitive personal and organizational data. The consequences of these attacks can be far-reaching, affecting not only individual organizations but entire economies and societies [5][6].

In response to these challenges, AI has emerged as a transformative force for cybersecurity. With its advanced capabilities in machine learning, natural language processing (NLP), and pattern recognition, AI has the potential to shift cybersecurity from a reactive to a proactive stance. It can detect known threats and predict and mitigate emerging threats efficiently. However, AI deployment in this domain is challenging. Critical questions surround the transparency of AI algorithms, the potential for inherent biases, and the ethical implications of automated decision-making in security contexts. Addressing these challenges is crucial for developing robust, transparent, ethical AI-driven cybersecurity frameworks. This presents a challenge and opportunity for significant innovation in the field [7].

The incorporation and application of AI technologies, such as attention mechanisms [8] and GANs [9], have particular significance in the context of cybersecurity. Attention mechanisms, which have gained recognition in NLP [10], exhibit remarkable effectiveness in enhancing the performance of neural networks by concentrating on the most pertinent aspects of data. This targeted attention facilitates the effective management of intricate datasets and the identification of subtle patterns, an invaluable capability in detecting refined cyber threats. In cybersecurity, attention mechanisms allow the identification of delicate and complex indicators of cyber threats often overlooked by conventional detection systems [11].

Similarly, GANs transform the generation of synthetic data [12]. Comprising a generator and a discriminator, these networks create data that resemble genuine scenarios. In cybersecurity, GANs tackle significant challenges, such as data scarcity and the need for diverse and representative attack scenarios. They enable the generation of innovative, realistic cyberattack scenarios, thereby bolstering the training and preparedness of cybersecurity models against a broad spectrum of threats, including novel and sophisticated attacks that are not represented in existing datasets [13].

The convergence of attention mechanisms and GANs presents a state-of-the-art approach to cybersecurity, significantly enhancing detection and response capabilities against a broad spectrum of cyber threats. This synergistic relationship bolsters the resilience and adaptability of cybersecurity systems and ensures their efficacy against the dynamic and sophisticated nature of current cyber threats. In an era marked by the escalating frequency, complexity, and impact of cyber threats, which results in substantial financial and reputational damage, the exploration and integration of advanced AI technologies, such as attention mechanisms and GANs, is not merely an academic pursuit but an essential measure. They represent a critical step towards developing proactive, adaptive, and resilient defense mechanisms capable of addressing the evolving challenges in cybersecurity.

The research questions posed earlier aim to guide the investigation into how these cutting-edge AI technologies can revolutionize cybersecurity strategies, rendering them more effective and resilient in an environment characterized by sophisticated and constantly evolving cyber threats. By focusing on integrating attention mechanisms for precise anomaly detection, employing GANs to simulate diverse cyber threats, and examining their combined impact on cybersecurity frameworks, this study contributes significantly to developing advanced, effective, and ethically sound cybersecurity solutions.

The primary achievements of this study are as follows:

1. Pioneering Fusion of AI Technologies: By merging attention mechanisms with GANs, this study introduces a novel approach in the realm of cybersecurity. This innovative fusion offers a more sophisticated and effective technique for recognizing and countering cyber threats and overcoming the constraints of conventional cybersecurity systems.

2. Enhanced Detection of Anomalies: The study demonstrates the potential of attention mechanisms to improve the accuracy and precision of anomaly detection in network traffic. This contributes to developing a more robust cybersecurity architecture that identifies intricate and subtle threats often overlooked by conventional methods.

3. Advances in Synthetic Data Generation: By applying GANs, this research provides valuable insights into creating realistic and diverse cyberattack scenarios. This addresses the pressing issue of data scarcity in cybersecurity, enabling a more comprehensive training and evaluation of cybersecurity models.

**Related Works**

Recent studies have highlighted the significance of artificial intelligence in devising proactive cybersecurity strategies. The AI Index Report 2023, published by Stanford University, offers extensive data on the progress made in AI technology, including its use in cybersecurity. The report underscores the growing trend of reliance on AI-driven security solutions [14]. A white paper by the Turing Institute examines the intricate AI in the cybersecurity terrain, utilizing a bottom-up and top-down methodology essential for comprehending AI's integration in this domain [15].

AI algorithms' ethical implications and transparency in cybersecurity are of significant concern. [16] delved into these issues, highlighting the need for ethical and transparent AI deployment in cybersecurity.

Recent studies have investigated the implementation of attention mechanisms in intrusion detection systems. Particularly, a study presented an efficient algorithm for intrusion detection using neural networks and attention architecture, illustrating the potential of attention mechanisms to improve cybersecurity [17].

GANs have gained considerable attention in cybersecurity, particularly owing to their ability to produce synthesized data that can imitate genuine scenarios. A survey dedicated to GANs in security explored their practical applications and the emerging challenges they present in device and network security [18].

The implementation of predictive AI in cybersecurity has garnered considerable interest. By employing predictive AI methods, cybersecurity systems can proactively anticipate forthcoming threats and demands, bolstering their capabilities [19].

The incorporation of AI in the realm of cybersecurity presents many challenges, including the requirement for substantial resources and ethical considerations. A discourse on these challenges offers valuable insights into overcoming the barriers to successfully integrating AI in cybersecurity [20].

The advancement and complexity of cyber threats have become pressing issues of great importance. Recent data and trends suggest that cyber threats are becoming increasingly sophisticated and relentless, presenting significant obstacles to conventional security measures [21].

The application of AI in anomaly detection, a vital component of cybersecurity, has been the subject of extensive research. AI-based systems are being developed to analyze and contrast data points within datasets to identify anomalies that could signal potential cyber threats [22]. Extensive analysis of cybersecurity in the context of the Internet of Things (IoT) has been conducted, providing a comprehensive examination of anomalies and security concepts within this domain [23].

This extensive review of the contemporary literature on AI and cybersecurity has revealed substantial progress and challenges in this area. Integrating AI technologies in cybersecurity is proving to be a transformative force; however, it also presents complex challenges that must be addressed. The research gaps identified in these studies provide opportunities for future innovation and development in creating more robust, efficient, and ethically sound cybersecurity solutions.

This study bridges several research gaps that have been identified in the literature:

1. Combined Application of AI Technologies: While attention mechanisms and GANs have individually demonstrated significant advancements in cybersecurity, research on their combined application is scarce. Our study pioneers this integration and offers a novel approach to countering cyber threats.
2. In-Depth Anomaly Detection Research: Although attention mechanisms show promise in anomaly detection, further research is required to evaluate their effectiveness in diverse and intricate real-world scenarios.
3. Advances in Synthetic Data Generation: The application of GANs to generate realistic cyberattack scenarios is an emerging field. Our study contributes to this area by providing insights into generating diverse and representative attack scenarios.

This study significantly advanced the development of state-of-the-art, efficient, and ethically sound cybersecurity solutions by addressing these gaps.

## Materials and methods

This section outlines our comprehensive methodology that integrates advanced machine learning techniques to enhance the detection and classification of network intrusions. The proposed method is illustrated in Figure 1 and detailed in Algorithm 1. Our approach primarily employs CNNs for effective feature extraction from sequential data. The CNN layers are adept at discerning intricate patterns within the data, making them particularly suited for analyzing complex network traffic and identifying potential anomalies. Moreover, we incorporate an attention mechanism to refine the feature extraction process further. This mechanism allows the model to focus on specific data segments that are more indicative of intrusion activities, thereby enhancing the overall accuracy and efficiency of the detection process.

To address the challenge of limited and imbalanced datasets, which can hinder the model's ability to simulate diverse intrusion scenarios, we employ a GAN. The GAN model is adeptly utilized to generate synthetic data, effectively augmenting our dataset with a wider range of intrusion scenarios. This augmentation not only enriches the training data but also assists in simulating more complex and varied intrusion scenarios, thereby substantially improving the robustness and reliability of our intrusion detection model.

The experiments used a 1.7 GHz Intel Core i7 processor, 16 GB of RAM, Windows 11 Professional 64-bit, and a 4 GB NVIDIA GeForce graphics card. Python and the TensorFlow library were used to develop the models.

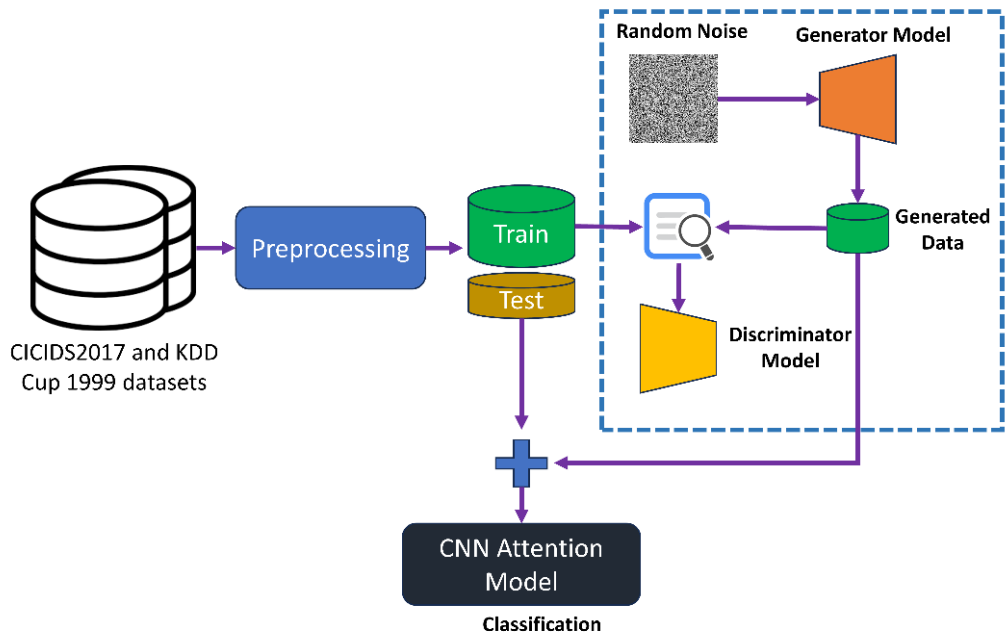

Figure 1: The architecture of the Attention-GAN model

Algorithm 1: The proposed method steps

| |
|---|
| **BEGIN** |
|     // Step 1: Data Preparation |
|     **LOAD** dataset |
|     **SPLIT** dataset into training set (X_train, y_train) and testing set (X_test, y_test) |
|     **ENCODE** categorical labels to numerical format |
|     RESHAPE X_train and X_test for CNN input requirements |
|     // Step 2: Train CNN with Attention Mechanism |
|     **DEFINE** CNN architecture with attention layers |

```
    COMPILE CNN model with appropriate loss function and optimizer
    TRAIN CNN model on training data (X_train, y_train)
    VALIDATE model on testing data (X_test, y_test)
    // Step 3: Train GAN
    DEFINE generator network in GAN that creates synthetic data
    DEFINE discriminator network in GAN that differentiates real from synthetic data
    COMPILE both generator and discriminator
    TRAIN GAN alternately on real data (to train discriminator) and on noise (to train
generator)
    // Step 4: Generate Synthetic Data
    GENERATE synthetic data using the trained generator of GAN
    RESHAPE synthetic data to match the shape of real training data
    // Step 5: Augment Data
    COMBINE real training data (X_train) with synthetic data
    // Step 6: Retrain CNN
    REDEFINE or REUSE CNN model architecture
    RETRAIN CNN model on combined dataset (real + synthetic)
    VALIDATE retrained model on testing data (X_test, y_test)
    // Step 7: Evaluate
    EVALUATE performance of the retrained model on testing data
    COMPARE performance metrics (accuracy, precision, recall, F1-score) before and after
augmentation
END
```

## 2.1 Dataset

In this project, we utilized the CICIDS2017 [24] and KDD Cup 1999 datasets [25]. The CICIDS2017 dataset is a comprehensive intrusion detection system (IDS) dataset featuring modern attack scenarios and network traffic patterns. On the other hand, the KDD Cup 1999 dataset is one of the most referenced datasets in network intrusion detection. It includes a wide range of simulated intrusions mixed with normal network traffic. Both datasets are integral for training and evaluating intrusion detection models, providing diverse scenarios to enhance the model's ability to detect various network intrusions.

In our cybersecurity model's data preparation and preprocessing phase, we undertook extensive measures to optimize the training conditions. Initially, we eliminated all duplicate entries to prevent model bias towards overrepresented records. Subsequently, we transformed categorical features into numerical values using a label encoder, enabling the machine learning algorithms to interpret them effectively. After that, we applied a MinMax scaler to normalize the numerical values, ensuring that each feature contributed equally to the model's training process without any single feature dominating due to scale. Finally, we divided our dataset into training and testing subsets with an 80:20 ratio to effectively develop and validate our model. This methodical preparation was essential in achieving a robust and dependable anomaly detection system.

## 2.2 Attention mechanism

The attention mechanism, a significant innovation in deep learning, emulates cognitive attention and is widely used in artificial intelligence, notably in natural language processing and computer vision. Originally developed for Seq2Seq models in neural machine translation, it addresses a key limitation of encoder-decoder models: their struggle to capture and retain information from long input sequences in a single fixed-length context vector.

The breakthrough by Bahdanau et al. [26]. It has involved the creation of adjustable shortcut connections between the context vector and the entire input, enhancing significant parts of the input while diminishing the rest. This mechanism is critical for handling longer inputs, as it prevents the model from forgetting parts of the data. It calculates multiple attention weights, denoted by $\alpha_{ij}$, to form the context vector $C_i$ for output $i$ as a weighted sum of annotations, using a softmax function:

$$C_i = \sum_{j=1}^{T_x} \alpha_{ij} h_j \tag{1}$$

$$\alpha_{ij} = \frac{exp(e_{ij})}{\sum_{k=1}^{T_x} exp(e_{ik})}, \tag{2}$$

$$e_{ij} = a(s_{i-1}, h_j) \tag{3}$$

Here, $e_{ij}$ is the output score from a feedforward network aiming to align input at $j$ with output at $i$.

GANs, introduced by Goodfellow et al., represent a pivotal advancement in unsupervised machine learning. They consist of two neural networks, the generator and the discriminator, which are trained simultaneously through a competitive process. The generator learns to produce data resembling the training set, while the discriminator learns to distinguish between real and generated data. As training progresses, the generator improves its ability to create realistic data, and the discriminator becomes better at identifying generated data. This training continues until the discriminator can no longer reliably differentiate between real and synthetic data, indicating the generator's proficiency in producing data indistinguishable from the real ones. Due to its ability to model complex data distributions, this model has found extensive applications in various fields like image generation, super-resolution, and more.

Our model uses a GAN to generate synthetic data for augmenting the training set. The GAN comprises two parts: a generator ($G$) and a discriminator ($D$). The generator aims to create indistinguishable data from real data, while the discriminator evaluates whether the given data is real (from the dataset) or fake (generated by $G$). The training process involves a min-max game, represented by the following equation:

$$min_G max_D V(D, G) = \mathbb{E}_{x \sim p_{data}(x)} \left[ \log D(x) \right] + \mathbb{E}_{z \sim p_z(z)} \left[ \log \left( 1 - D\left( G(z) \right) \right) \right] \tag{4}$$

Here, $p_{data}(x)$ is the data distribution, $G(z)$ generates data from the noise distribution $p_z(z)$, and $D(x)$ outputs the probability that $x$ came from the real data rather than $G$. The generator improves data generation capability while the discriminator enhances detection accuracy, creating realistic synthetic data.

## 2.3 Training process

The training process for the CNN with an attention mechanism and the GAN model is detailed and involves several layers and parameters:

CNN with Attention Mechanism (Table 1 and Table 2 show the model architecture and parameters for the CICIDS dataset and KDD dataset, respectively) :

- Architecture: The model starts with an input layer shaped to (sequence_length, input_dim). It includes two Conv1D layers, the first with 32 filters and the second with 64, using a kernel size of 3, 'same' padding, and ReLU activation. Following the convolutional layers is a global average pooling layer.
- Attention Layer: An attention layer follows, using a reshape layer to adapt its input.
- Output Section: A dense layer with 128 units and ReLU activation leads into a dropout layer with a 0.5 rate, culminating in a final dense layer with sigmoid activation for output.
- Training Parameters: Compiled with the Adam optimizer and binary cross-entropy loss, the model trains over 10 epochs with a batch size 128.

Table 1: Model architecture and parameters of CNN with Attention Mechanism for CICID dataset

| Layer (type) | Output Shape | Param No. |
|---|---|---|
| Input Layer | (None, 78, 1) | 0 |
| Conv1D | (None, 78, 32) | 128 |
| Conv1D | (None, 78, 64) | 6208 |
| GlobalAveragePooling1D | (None, 64) | 0 |
| Reshape | (None, 1, 64) | 0 |
| Attention | (None, 1, 64) | 1 |
| Dense | (None, 1, 128) | 8320 |
| Dropout | (None, 1, 128) | 0 |
| Dense | (None, 1, 1) | 129 |

Table 2: Model architecture and parameters of CNN with Attention Mechanism for KDD dataset

| Layer (type) | Output Shape | Param No. |
|---|---|---|
| Input Layer | (None, 30, 1) | 0 |
| Conv1D | (None, 30, 32) | 128 |
| Conv1D | (None, 30, 64) | 6208 |
| GlobalAveragePooling1D | (None, 64) | 0 |
| Reshape | (None, 1, 64) | 0 |
| Attention | (None, 1, 64) | 1 |
| Dense | (None, 1, 128) | 8320 |
| Dropout | (None, 1, 128) | 0 |
| Dense | (None, 1, 5) | 645 |

a. GAN model:

The GAN model's generator commences with a dense layer comprised of 30 units and employs a LeakyReLU [27] activation function with an alpha of 0.01 to introduce non-linearity while mitigating the issue of vanishing gradients. The layer output is reshaped to form a sequence that matches the time steps of the input. The discriminator features a two-tier Conv1D structure, with the first layer comprising 64 kernels and the second 32, employing LeakyReLU activation and a stride of 2 for extracting features at varying scales. A flatten layer follows to convert the convolved features into a vector, which then feeds into a Dense layer with a sigmoid activation, producing a probabilistic assessment of whether the input sequence is real or generated. The model trains through an iterative adversarial process across several epochs, optimizing the generator to create indistinguishable data from real data and refining the discriminator's ability to classify the sequences accurately. Combining these two models, with their respective architectures and training parameters, is crucial to enhancing the system's capability in accurately detecting network intrusions, leveraging the CNN's feature extraction prowess and the GAN's data augmentation strength.

**2.4 Model Evaluation and Performance Metrics:**

After completing the training, the performance of the Deep Learning model was evaluated using a series of standard metrics for classification problems, including accuracy, precision, recall, F1-score, and the confusion matrix. These metrics provide distinct viewpoints on the model's performance and are constructed based on the concepts of true positives (TP), true negatives (TN), false positives (FP), and false negatives (FN). These values are often depicted in a confusion matrix, a tabular layout that allows for visualizing the algorithm's performance.

2.4.1 Accuracy
Accuracy measures the model's overall ability to predict both classes correctly, represented as the ratio of correctly predicted observations (both positive and negative) to the total number of observations. However, it may be misleading when the classes are imbalanced. The formula for calculating accuracy is as follows:

$$Accuracy = \frac{(TP+TN)}{(TP+FP+FN+TN)} \qquad (5)$$

2.4.2 Precision
Precision, or Positive Predictive Value (PPV), is a metric that assesses the proportion of accurate positive identifications among all positive identifications made. It is particularly crucial when the cost of a false positive is substantial. Precision is calculated by dividing the number of true positive identifications by the total number of positive identifications made.

$$Precision = \frac{TP}{(TP+FP)} \qquad (6)$$

2.4.3 Recall
Recall, or Sensitivity, measures the proportion of actual positives that were identified correctly. It is especially important when the cost of a false negative is high. It is calculated as follows:

$$Recall = \frac{TP}{(TP+FN)} \hspace{2cm} (7)$$

### 2.4.4 F1-score

The F1-score is a quantitative measure that incorporates precision and recall, with equal emphasis on each metric. It obtains its highest value of 1 when precision and recall are perfect and its lowest value of 0 when absent. This metric is particularly useful in instances where the class distribution is imbalanced. The F1-score is determined by calculating the harmonic mean of precision and recall.

$$F1 - score = 2 * \frac{Precision * Recall}{(Precision + Recall)} \hspace{1cm} (8)$$

## 3 Results and discussion

### 3.1 CICIDS2017

The results from the proposed method on the CICIDS2017 dataset, as shown in Table 3, indicate a consistent improvement in accuracy and loss metrics over ten epochs. Starting with an accuracy of 92.92% and a loss of 0.1781 in the first epoch, the model demonstrates progressive enhancement, reaching an accuracy of 97.93% and a loss of 0.0448 by the tenth epoch. Similarly, the validation accuracy improves from 95.47% to 97.90%, with a corresponding decrease in validation loss. The confusion matrix further substantiates the model's effectiveness with high true positive (TP) and true negative (TN) rates and relatively lower false positive (FP) and false negative (FN) rates. The calculated precision, recall, and F1-scores for both classes are above 97%, indicating a high level of model reliability and balanced performance across different classifications. This performance suggests the model's robustness in accurately detecting and classifying network intrusions.

Upon examining the CICIDS dataset, it becomes evident that the model exhibits impressive predictive capabilities, as evidenced by the confusion matrix and accuracy graphs provided. The training accuracy commences strongly and continues to increase steadily, while the validation accuracy tracks closely, suggesting that the model is acquiring generalizable features. The loss graphs reveal a consistent decline in training and validation loss, indicating a stable learning process across the epochs.

The confusion matrix in Figure 2 indicates a high true positive rate, which signifies accurate attack detection. The true negative rate is also substantial, reflecting the model's precision in distinguishing normal behavior. The low false positive and false negative rates further confirm the model's efficacy.

The precision, recall, and F1-scores suggest that the model performs with balance across classes, with values demonstrating high precision in differentiating between various types of network traffic. These outcomes are highly encouraging, reflecting the model's potential as a reliable tool for network intrusion detection in cybersecurity.

The training curves for the CICIDS2017 dataset (Figure 4 [a, and b]) illustrate a successful model optimization. The loss curve displays a sharp decline in the initial epochs, leveling off as the model converges, indicative of an effective learning process. Similarly, the accuracy curve shows a steep ascent in the initial training phase, reaching a high plateau, which suggests the model's high predictive capability. The training and validation curves are closely aligned, reflecting the model's generalization ability without significant overfitting. This congruence between training and validation performance underscores the robustness of the model.

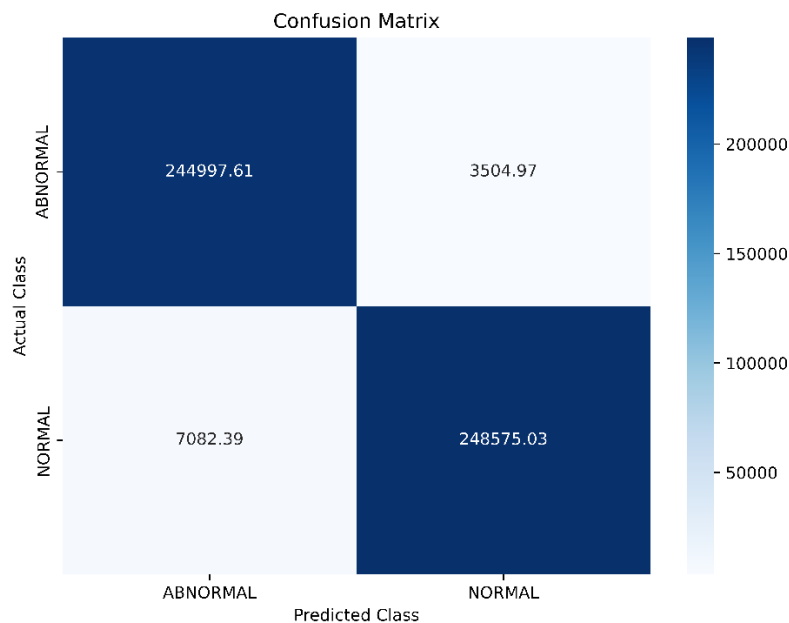

Figure 2: The confusion matrix for the test data on the CICIDS2017 dataset.

### 3.2 KDD dataset

The results from the proposed method on the KDD dataset over 10 epochs show significant improvement. The model began with an accuracy of 94.61% and a loss of 0.1843, rapidly progressing to an impressive 99.69% accuracy with a loss of just 0.0116 by the final epoch. This demonstrates the model's efficiency in learning and adapting to the dataset's characteristics. The validation accuracy also shows a similar positive trend, reaching 99.76%. The precision, recall, and F1-scores across different classes indicate high accuracy, especially for class 0. The model's slightly lower performance in classes 3 and 4 (with lower recall and F1-scores) might indicate a need for further tuning or more data for these classes. These results highlight the model's effectiveness in intrusion detection tasks using the KDD dataset.

The model's performance on the KDD dataset has been depicted through loss and accuracy diagrams (Figure 4 [c, and d]), demonstrating its efficient learning and strong generalization capabilities. The rapid decrease in loss and simultaneous rise in accuracy during the initial epochs highlight a robust learning process. The convergence of both metrics signifies the attainment of optimal performance. The close correspondence between the training and validation lines in both diagrams highlights the model's stability, suggesting that it has not overfitted and retains high predictive power for unseen data. This equilibrium is a hallmark of a well-tuned model suitable for deployment in detecting network intrusions.

 The confusion matrix (Figure 3) derived from the evaluation of the KDD dataset affords a substantial quantity of true positives and true negatives, thereby exemplifying the model's aptitude in accurately classifying both normal and attack instances. The low false positive and false negative rates corroborate the model's potent discriminative capabilities, with minimal occurrences of misclassification. These findings are additionally substantiated by the precision, recall, and F1-score metrics, which illustrate exceptional model performance across a diverse range of classes. However, there is a discernible need for improvement in accurately classifying certain minority classes (0,1,2,3 and 4 represent normal, DoS, U2R, Probe, and R2L, respectively).

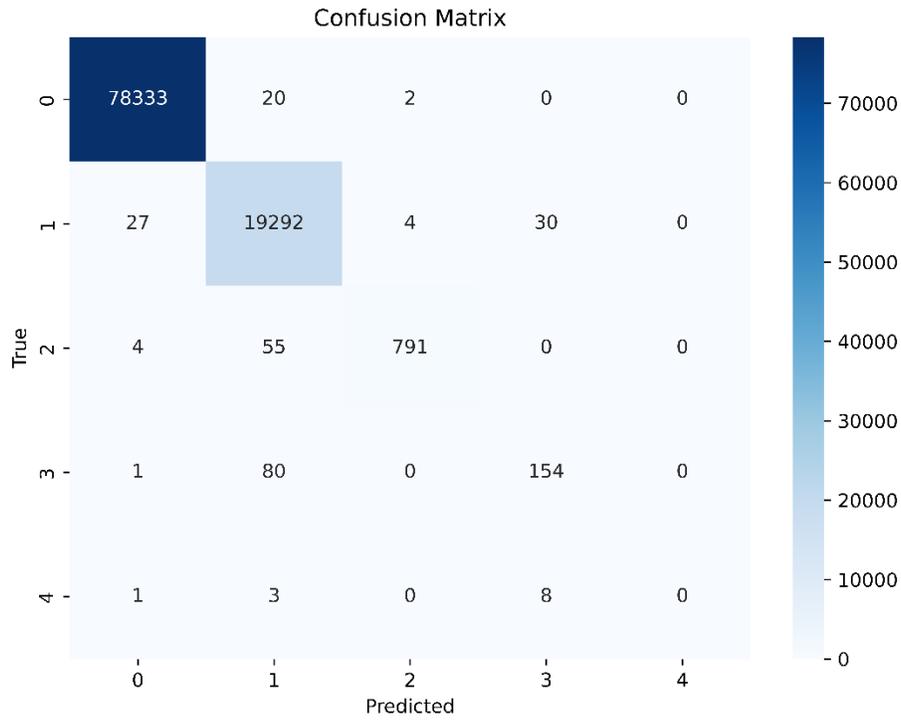

Figure 3: The confusion matrix for the test data on KDD dataset.

Table 3: The performance metrics comparison between the KDD and CICIDS datasets.

| Metric | KDD Dataset (Macro Avg) | KDD Dataset (Weighted Avg) | CICIDS Dataset (Class 0) | CICIDS Dataset (Class 1) |
|---|---|---|---|---|
| Precision | 0.76 | 1.00 | 0.9859 | 0.9723 |
| Recall | 0.72 | 1.00 | 0.9719 | 0.9861 |
| F1-Score | 0.74 | 1.00 | 0.9788 | 0.9791 |

Table 3 compares the precision, recall, and F1-score across both datasets. The macro average provides a view of the overall performance across all classes, while the weighted average accounts for the support of each class. The CICIDS dataset shows remarkably high precision, recall, and F1-scores, indicating excellent model performance across both classes.

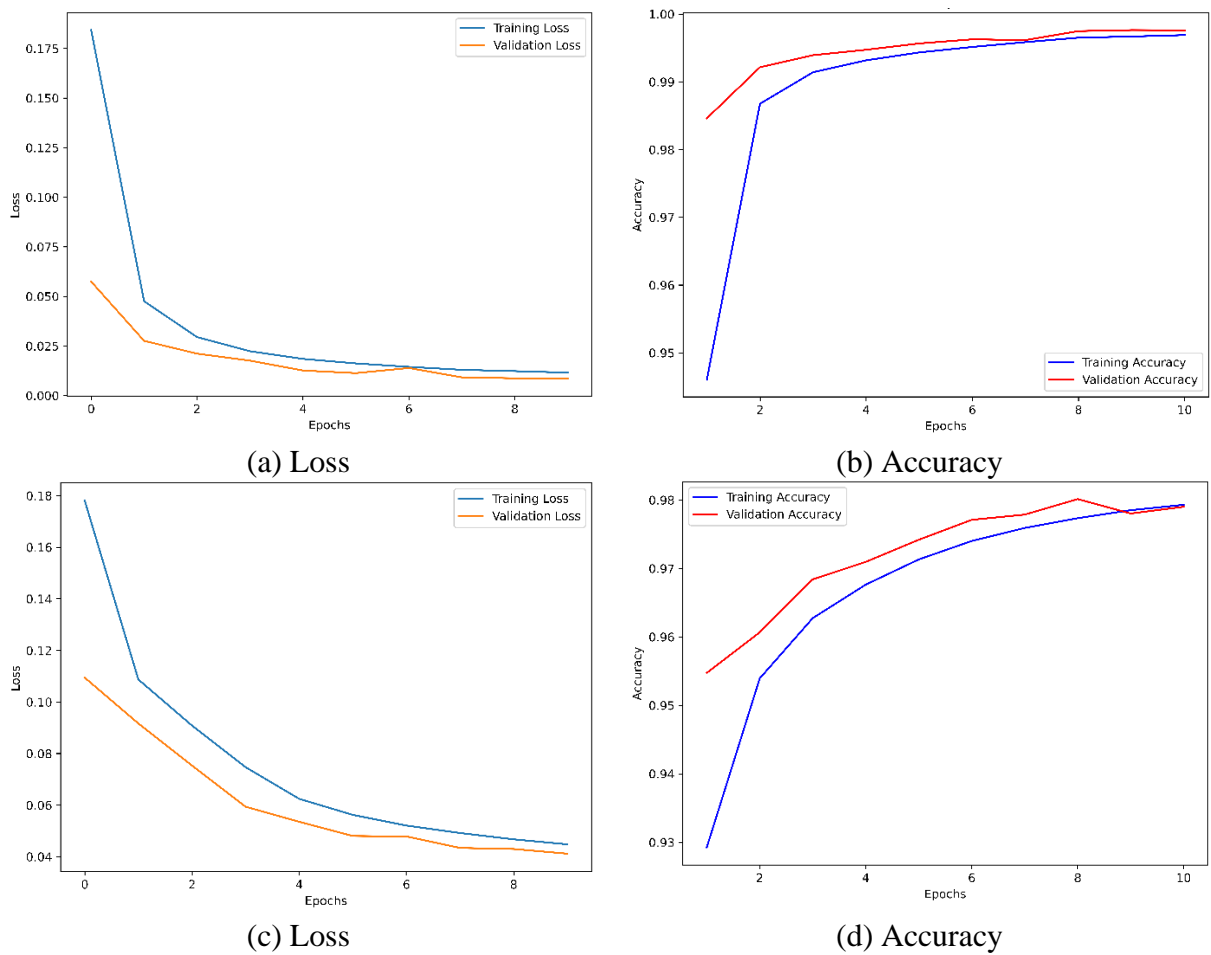

(a) Loss          (b) Accuracy

(c) Loss          (d) Accuracy

Figure 4: The training plots for the proposed model a and b for the CICIDS2017 dataset, c and d for the KDD dataset.

## 3.3 Comparison with the state-of-the-art methods

Assessing emerging deep learning methodologies considering state-of-the-art approaches, is paramount in cybersecurity. The recognition of progress facilitated by innovative approaches necessitates this contextualization. Conventional web-based threat prevention tactics include anomaly and signature detection in conjunction with machine learning techniques such as decision trees, support vector machines, and ensembles. The domain of deep learning has experienced considerable advancements in recent years, with a particular focus on recurrent neural networks (RNNs) and convolutional neural networks (CNNs). However, this emphasis has frequently overlooked the potential of transformer models. Our proposed models are evaluated concerning current best practices as documented in existing academic literature, as presented in Table 4. This evaluation is based on F1 scores on the CICIDS2017 dataset.

Table 4: Comparison with state-of-the-art methods.

| Reference | F1-Score |
|---|---|
| [28] | 0.90 |
| [29] | 0.89 |
| [30] | 0.88 |
| [31] | 0.86 |
| [32] | 0.85 |
| [33] | 0.94 |
| Attention-GAN (proposed method) | >0.97 |

**4 Conclusion**

The present research, which incorporates a CNN with an attention mechanism and a GAN on the CICIDS2017 and KDD datasets, has substantially advanced network intrusion detection. The notable accuracy and minimal loss rates attest to these models' effectiveness in identifying intricate intrusion patterns. However, areas of improvement include addressing the variation in performance among different classes, particularly those that are underrepresented, by enhancing data representation and model tuning. Future studies should focus on overcoming class imbalance and refining models to improve detection in minority classes. Furthermore, exploring more advanced neural network architectures and diversifying the training data could lead to even more reliable and precise intrusion detection systems.